\documentclass{article}
\usepackage{ijcai24}

\usepackage{times}
\usepackage{soul}
\usepackage{url}
\usepackage[hidelinks]{hyperref}
\usepackage[utf8]{inputenc}
\usepackage[small]{caption}
\usepackage{graphicx}
\usepackage{amsmath}
\usepackage{amsthm}
\usepackage{booktabs}
\usepackage{algorithm}
\usepackage{algorithmic}
\usepackage[switch]{lineno}
\usepackage{makecell}
\usepackage{multirow}
\usepackage{array}
\usepackage{bbding}
\usepackage {setspace}

\title{
Affective Computing for Healthcare: Recent Trends, Applications, Challenges, and Beyond
}

\author{
Yuanyuan Liu$^1$\and
Ke Wang$^1$\and
Lin Wei$^{1}$\and
Jingying Chen$^2$\and
Yibing Zhan$^3$\\
Dapeng Tao$^4$\And  
Zhe Chen$^5$\\  
\affiliations
$^1$School of Computer Science, China University of Geosciences, Wuhan, China\\
$^2$Central China Normal University, China\\
$^3$JD Explore Academy, China\And
$^4$Yunnan University, China\\
$^5$The School of Computing, Engineering and Mathematical Sciences, La Trobe University, Australia.\\
}

\begin{document}

\maketitle

\begin{abstract}

Affective computing, which aims to recognize, interpret, and understand human emotions, provides benefits in healthcare, such as improving patient care and enhancing doctor-patient communication. However, there is a noticeable absence of a comprehensive summary of recent advancements in affective computing for healthcare, which could pose difficulties for researchers entering this field. To address this, our paper aims to provide an extensive literature review of related studies published in the last five years. We begin by analyzing trends, benefits, and limitations of recent datasets and affective computing methods devised for healthcare. Subsequently, we highlight several healthcare application hotspots of current technologies that could be promising for real-world deployment. Through our analysis, we identify and discuss some ongoing challenges in the field as evidenced by the literature. Concluding with a thorough review, we further offer potential future research directions and hope our findings and insights could guide related researchers to make better contributions to the evolution of affective computing in healthcare.

\end{abstract}

\begin{figure*}[t]
   \centering
   \includegraphics[width=1.0\textwidth]{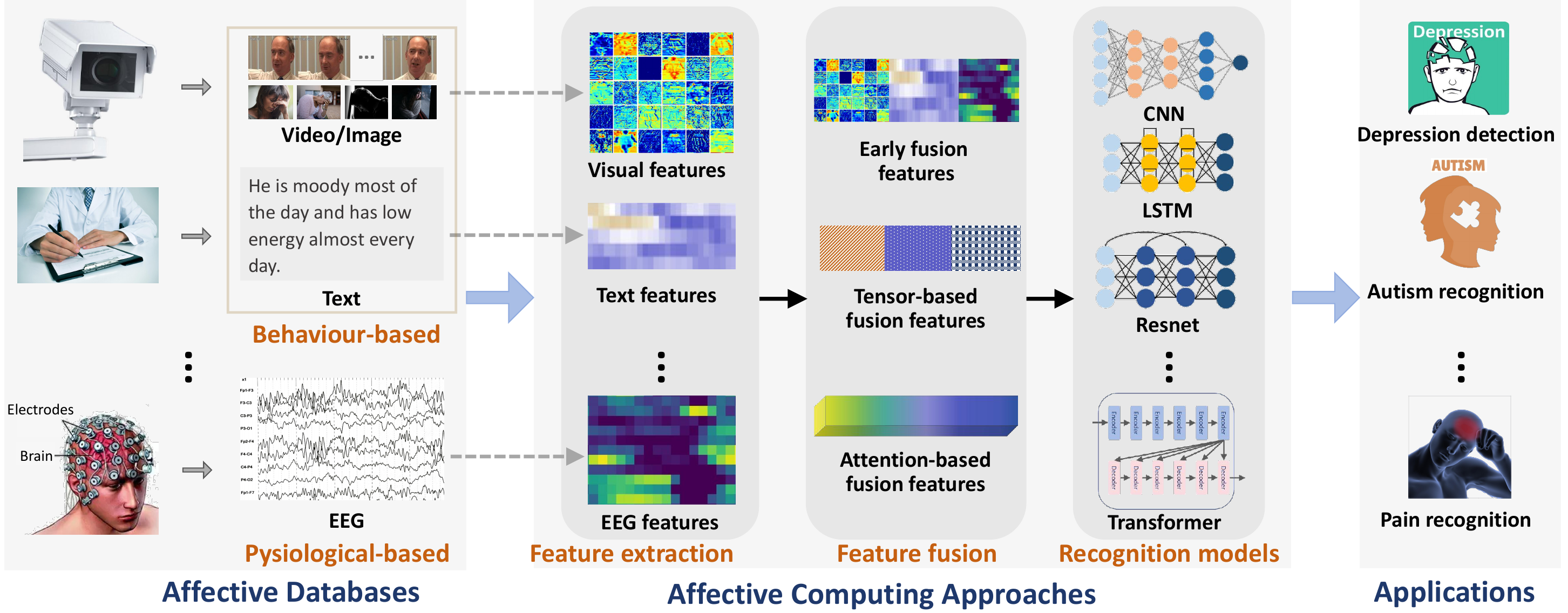}
   \caption{Framework of learning-based affective computing for healthcare.}
   \label{Fig.framework}
\end{figure*}

\section{Introduction}
Affective computing refers to the process of utilizing computer and artificial intelligent technologies to analyze and recognize human emotions
\cite{hossain2019emotion}. 
Human emotions, the intuitive reflections of the body's state, can convey useful information for treatment processes. By accurately identifying patient emotions 
through affective computing, 
doctors can obtain more informative patient statistics and provide more appropriate or customized treatments for patients. 
As a result, affective computing could play a pivotal role in aiding doctors with various tasks like mental health monitoring, personalized treatment, and patient support. Due to promising applications, affective computing for healthcare has attracted increasing attention \cite{alelaiwi2019multimodal,1,2}. 

In particular, thanks to the significant modeling capabilities brought by deep learning advancements, affective computing has shown outstanding capacities of representing and identifying various human-related information \cite{rejaibi2019clinical,smith2020vocal}. Diverse deep learning models, such as Convolutional Neural Networks (CNN), Recurrent Neural Network (RNN), and Transformer, have been introduced to extract emotional features from either unimodal or multimodal data, resulting in accurate emotion recognition and analysis. 
Meanwhile, breakthroughs in affective computing have also led to impressive progress in complex healthcare scenarios and applications. 
Despite promising progress, we found that existing related reviews mostly summarise the affective computing techniques in a relatively narrow application area like depression recognizing \cite{giuntini2020review}. 
We believe that the lack of a comprehensive and insightful review of affective computing in healthcare may create unnecessary barriers for researchers to enter/understand the field.

To bridge the gap mentioned above and summarize the progress achieved in recent deep learning-related advancements, we perform a comprehensive literature review of the recent progress in affective computing for healthcare in this paper. An overview of our work is shown in Figure \ref{Fig.framework}. In general, we attempt to thoroughly review the recently published datasets and approaches, covering a broad area related to affective computing for healthcare. The contributions of this paper and differences from other surveys are as follows: 

\textbf{1)} We identify 3 major research directions based on methodologies, including behavior-based, physiological-based, and behavior-physiological-based research. Under this taxonomy, we attempt to perform a comprehensive analysis of the recent development trends of datasets and approaches, illustrating their details, benefits, as well as their limitations. 

\textbf{2)} We highlight a few most frequently focused medical application directions of affective computing, \emph{i.e.}, depression diagnosis \cite{othmani2021towards}, autism recognition and intervention \cite{li2021two}, pain level recognition \cite{phan2023pain}, and other related medical applications \cite{ayata2020emotion}, establishing synergies between affective computing and clinical applications.  

\textbf{3)} In addition to existing achievements, we present significant challenges that still pose difficulties for developing and applying affective computing in healthcare. Despite the challenges, we also conclude potential directions for future research, hoping to provide valuable insights for researchers who would like to contribute to this field. 

\begin{table*}[t]
    \centering
    \scalebox{0.72}{
    \begin{tabular}{ccccccc}
       \toprule[2pt]
        
        Class &Datasets  & \#Subjects & \#Samples  & Modalities & Applications  & Labels  \\
       
        \hline
        \multirow{9}*{Behaviour}

        &\multicolumn{1}{l}{FENP  ~\cite{yan2020fenp}} & 106 &11000 & Visual & Pain &  Pain, No Pain   \\
        &\multicolumn{1}{l}{PersionSIChASD  ~\cite{Alizadeh2021APS}} & 38 &418 &  Audio & Autism & Autistic, Normal, Phonetic units   \\
        &\multicolumn{1}{l}{D4 ~\cite{yao2022d4}}  & 201 &1339 & Text &   Depression  & Risk, Non Risk \\
         &\multicolumn{1}{l}{MMDA  ~\cite{jiang2022mmda} } & 962 & 1025 &Text, Audio, Visual& Depre,Anx & Depre,Nondepre,Anx,Nonanx   \\
       & \multicolumn{1}{l}{D-Vlog ~\cite{yoon2022d}}  & 816 &961  &  Audio, Visual & Depression & Depre, Nondepre   \\
       
        &\multicolumn{1}{l}{SAD  ~\cite{alghifari2023development}}  & 64 &64 & Audio &  Depression  & Depre, Nondepre    \\

        &\multicolumn{1}{l}{TIAD  ~\cite{melinda2023novel}} & 34& 6120 &Visual &  Autism  &  Autistic, Normal \\	
       
        &\multicolumn{1}{l}{PEMF  ~\cite{fernandes2023pain}}  & 68 & 272&Visual&   Pain&  Pain, No Pain    \\
        &\multicolumn{1}{l}{ CALMED  ~\cite{sousa2023introducing}} &  4 &57012&  Audio, Visual & Autism & Autistic, Normal  \\
        
        \hline
        \multirow{2}*{Physiological}
        &\multicolumn{1}{l}{DDLES  ~\cite{8786540}} & 60&60  &EEG& Depression&Minimal,Mild,Moderate,Severe  \\
         &\multicolumn{1}{l}{SADR~\cite{8653893}}  & 39& 1170  &EEG,EM& Depression & Depre, Nondepre  \\
        \hline
        \multirow{3}*{Behaviour}
        &\multicolumn{1}{l}{EMCASD  ~\cite{emcasd} } & 28& 300  &Visual, Other & Autism  & Autistic, Normal  \\
        &\multicolumn{1}{l}{MMOD  ~\cite{Cai2020AMO} } & 160& 160  &Audio, EEG& Depression  & Depre, Nondepre  \\
        -Physiological &\multicolumn{1}{l}{PPAD  ~\cite{salekin2021multimodal} } & 58  &58 &Audio, Visual, Other& Pain  & Pain, No Pain  \\
        &\multicolumn{1}{l}{M-Ms  ~\cite{calabro2021m} } & 15 &  37127 & Visual, EEG&  Autism & Good, Poor   \\
         
        \bottomrule[2pt]
    \end{tabular}}
    \caption{The most recent affective computing datasets in the medicine field. Depre, Nondepre, Anx, and Nonanx are abbreviations for depression, nondepression, anxiety, and nonanxiety, respectively.}
    \label{tab: plain}
\end{table*}

\section{Datasets and Approaches}

We first perform a literature review on affective computing research, mainly covering the related datasets and approaches applied in healthcare in the last five years. Tables \ref{tab: plain} and \ref{tab: plain1} list the overview of databases and typical approaches, respectively. We find that affective computing research varies according to the type of data utilized. Hence, we classify related research into three categories: behavior data-based,  physiological data-based, and behavior-physiological data-based research. Each category of research is detailed below.

\subsection{Behavior Data-based Research}
In the realm of healthcare, behavior data-based research entails employing affective computing technology to analyze patients' emotions through their behavior data, such as language, facial expressions, and medical records.  
\paragraph{Datasets}  
Behavioral affective datasets can be discussed based on three categories: vision-based datasets, non-vision-based unimodal datasets, and multimodal datasets, depending on their popularity and diverse modalities. 

\underline{Vision-based datasets} are the mainstream dataset format, partly due to the widespread applications of vision techniques. Current visual affective databases are generally datasets with patients' facial expressions. TIAD collected thermal imaging of the faces of autistic children~\cite{melinda2023novel}, FEMP collected 11,000 facial images of newborns~\cite{yan2020fenp}, and PEMF consists of 272 micro-clips with facial images~\cite{fernandes2023pain}. Through the analysis of facial expressions, important emotional clues can be obtained to help doctors judge the emotional state of patients and improve the diagnosis of disease. 

\underline{Non-vision unimodal datasets} contain text or audio data modalities. \emph{Text}: D4 records the conversations between doctors and patients during depression diagnosis. The diagnosis results and symptom summaries given in each conversation~\cite{yao2022d4}. \emph{Audio}: SAD is an audio dataset of English depression that contains 64 recordings of individuals with and without depression~\cite{alghifari2023development}. 

\underline{Multimodal datasets} primarily combine two or three modalities of text, audio, and visual data, providing richer cues to patients' emotional behaviour, considering the emotional bias inherent in single behavioral data.  
D-vlog comprises 961 vlogs collected from YouTube that encompass both audio and videos ~\cite{yoon2022d}. CALMED consists of audio and video features extracted from conference transcript files of children diagnosed with autism~\cite{sousa2023introducing}. MMDA is the largest mental disorder dataset including visual, auditory, and textual data, where all subjects in MMDA are diagnosed by professional psychologists using de-identified original interview videos~\cite{jiang2022mmda}. 

\paragraph{Approaches}  
As shown in Table \ref{tab: plain1}, deep learning models have become the dominant approaches for affective computing in recent years and are gradually evolving from unimodal approaches towards multimodal approaches.

\underline{Vision-based affective approaches} primarily employ CNN and attention methods to 
extract emotional features from facial images and videos, identifying emotion categories or scores.  
Typical methods include: ~\cite{liu2023net} built a facial expression recognition model based on the self-attention mechanism  for depression detection.
\cite{lu2023video} used a two-stream CNN network with cross-stream attention mechanism to integrate spatial and temporal information in newborn facial expression videos for pain level recognition.

\underline{Text-based unimodal affective approaches} use natural language processing technology to analyze text data to identify emotions. Typical methods are as follows: \cite{dessai2022depression} used CNN and Long Short Term Memory (LSTM) to analyze Twitter users' tweets to discover factors related to depression. In \cite{ji2023towards}, the authors utilized Large Language Models’ (LLM) interactivity and multitasking ability to investigate ``hallucination'' in medical question-answering systems. That is, the model produces information that sounds reasonable but is not faithful or meaningless.

\underline{Audio-based unimodal affective approaches} usually adopt CNN and LSTM to capture the long-term dependence and high-level affective feature representation of audio sequences. 
\cite{han2023spatial} developed a self-supervised learning framework that combines integrate causal and dilated convolution to continuously enlarge the receptive domain for capturing multi-scale emotion-contextual information, and employed a hierarchical contrast loss to predict depression by exploring the long temporal emotion dependencies of audio. 

 \underline{Multimodal affective approaches} have emerged as a trend in healthcare affective computing. This approach combines visual, audio, and text signals to more accurately identify patients' emotions, surpassing the results of unimodal affective analysis.  
For example, \cite{li2021two} used pre-trained ResNet-18 and ResNet-50 to construct a two-stage network for children autism prediction from facial expression videos and audio.
\cite{wang2023multimodal} first used the pre-trained models ALBERT and AGG16 to construct a visual-language model to extract facial expression image features, text, and behavioral features, and then used the early fusion method to fuse these features to form multimodal features for depression degree detection.
\cite{zhao2022unaligned} proposed a cross-modal attention mechanism-based Generative Adversarial Network, and used an attention-based fusion approach to integrate facial expression videos, text, and audio for automatically depression severity assessment.


\subsection{Physiological Data-based Research}
Physiological data-based affective computing research focuses on 
integrating physiological signals into computational models for robust sentiment analysis. Physiological signals record and measure an individual's affective state, affective experience, or affective response, such as Electroencephalography (EEG), offering a more objective reflection of the genuine emotional state \cite{BAJESTANI2019277}.

\paragraph{Datasets} As shown in  Table \ref{tab: plain},  physiological affective datasets mainly contain emotion-related physiological signals, \textit{i.e.}, EEG,  Electrodermal activity (EDA), and other related signals (\textit{e.g.}, Electrocardiogram (ECG), Electromyography (EMG), Heart Rate (HR), Eye-tracking (ET), Accelerometer (ACC), Inter-Beat Interval (IBI), Blood Volume Pulse (BVP), and Skin Temperature (ST)). They also can be divided into \underline{unimodal physiological datasets} and \underline{multimodal physiological datesets}.
DDLES, an EEG-based depression recognition dataset, is a typical unimodal physiological dataset that contains EEG signals of 60 depressed and non-depressed individuals~\cite{8786540}.
SRDR is a multimodal physiological dataset that monitors specific physiological variables and synchronously collects EEG and EM signals from subjects to provide a more accurate detection dataset in a clinical environment~\cite{8653893}.
Since the acquisition of physiological signals requires expensive specialised medical equipment, most of the existing relevant datasets are small. 
 
\paragraph{Approaches}  
Physiological-based affective analysis methods usually use CNN and RNN to process temporal physiological data, which can extract the dynamic time-frequency features of physiological signals, facilitating a better understanding of dynamic affective states. Depending on the signal data, these methods also can be classified into two categories: {unimodal approaches} and {multimodal approaches}.
\underline{Uimodal approaches} focus on mining emotional states from one type of emotional physiological signal. \cite{xia2023end} is a typical unimodal approach that uses multi-head self-attention mechanism and double-branch CNN to construct an end-to-end model for depression recognition from single EEG signals. 
In contrast, \underline{multimodal approaches} explore multiple types of physiological signals in the emotion recognition process, thus improving the prediction accuracy. 
 \cite{phan2023pain} first used CNN and BiLSTM to extract the low-level features and time information in the sequence from ECG and EDA signals, and then combined ECG and EDA features in an early fusion manner for pain recognition.

\subsection{Behavior-Physiological Data-based Research}
Due to the directness and interpretability of behavioural data and the high objectivity of physiological data, the research integrating both behavior and psychological data is a natural derivation for the application in healthcare.

\paragraph{Datasets}  
Behavior-Physiological datasets integrate physiological data like EEG, EDA, and EMG with  behavioral data such as facial expressions and audio. This integration aims to capture more comprehensive emotion-related cues of patients. 
M-MS is a multimodal autism emotion dataset, containing ECG and therapy videos to support the study of synchronization in autism recognition~\cite{calabro2021m}. 
PPAD is the first multimodal neonatal pain dataset containing facial expression videos, sound, and physiological responses (vital signs and cortical activity), which was collected from 58 neonates during their hospitalization in the neonatal intensive care unit~\cite{salekin2021multimodal}. 
 
\paragraph{Approaches}  
Behavior-Physiological affective analysis needs to mine both the external manifestations of behavioural data and the internal changes of physiological signals, aiming to achieve more accurate affective analysis performance. 
For example, \cite{han2022multimodal} proposed a multimodal diagnostic framework that uses an early fusion approach combining EEG and ET data for unsupervised training and supervised fine-tuning to identify children with Autism Spectrum Disorder (ASD).
Recently, \cite{qayyum2023high} employed ViT, CNN, and LSTM models for the extraction of audio and EEG features. Subsequently, these features were concatenated by early fusion to improve the diagnostic performance of depression.

\subsection{Discussion}
  


 \paragraph{Tendency}
 In general, in the healthcare field, in contrast to unimodal technology, there is a growing research trend towards multimodal affective computing technology that analyzes both behavior and physiological signals. This allows for the exploration of both external manifestations of behavioral data and internal changes in physiological signals.
 Moreover, affective computing approaches have also shifted from early CNN-based models to Transfomer-based models, and then to hybrid large-scale ones, which can better focus on sentiment changes across various modality sequences. 
Both facilitate a more comprehensive analysis of a patient's emotional state, assisting doctors in tasks such as mental health monitoring, personalized treatment, and patient support. 
 
 \paragraph{Limitation}
Despite the progress made, we find that there are still limitations in the current technology, mainly including: \textbf{For datasets}:  Behavioural datasets are easy to collect but subjective and deceptive, and physiological datasets are highly reliable but costly and difficult to collect.  As a result, existing multimodal fusion datasets remain limited in size, hindering their ability to fully support complex healthcare and medical applications; \textbf{For approaches}: As shown in Table \ref{tab: plain1}, although higher recognition rates are achieved in current studies, most existing affective models are trained and designed independently on smaller and more limited set datasets. 
This lack of interoperability makes it challenging to reuse these models across different healthcare systems and applications. 
\begin{table*}
    \centering
     \scalebox{0.67}{
    \begin{tabular}{cccccccccccc}
          \toprule[2pt]
        \multirow{2}*{Class} &  \multirow{2}*{Methods} &  \multirow{2}*{Technology}  & \multirow{2}*{Modalities} &  \multirow{2}*{Application}&\multirow{2}*{Dateset} & \multicolumn{5}{c}{Performance}\\
        \cmidrule(r){7-11}
        &&&&& &Acc.& F1 & Pre. & Rec.&MAE\\
        
        \midrule
        \multirow{15}*{B}
         
          &    \multicolumn{1}{l}{\cite{Rejaibi2019ClinicalDA} }  &CNN   &  Audio  &  Depression &DAIC-WOZ&73.25 &-&-&-&-\\
          
        &    \multicolumn{1}{l}{\cite{peng2020pain} } &Multi-Scale DNN   &  Visual  &  Pain &UNBC Shoulder Pain&79.94 &-&-&-&0.57\\
        
        
        &    \multicolumn{1}{l}{\cite{9320205} }   &RNN-LSTM &  Visual  &  Pain &EmoPain &- &-&-&-&-\\
        
        &   \multicolumn{1}{l}{\cite{li2021two}} &ResNet  &  Audio, Visual  &  Autism &ASD-Affect&72.40&75.00&-&-&- \\
        
        &    \multicolumn{1}{l}{\cite{9688837} } &CNN-LSTM-Transformer  & Audio,Text &Depression &Own&66.00 &-&-&-&-\\
        &   \multicolumn{1}{l}{ \cite{dessai2022depression}} &CNN-LSTM & Text  &Depression &Own&92.00&93.00&93.00&94.00&- \\
         &    \multicolumn{1}{l}{\cite{mohan2022edge}} &DCNN   &  Visual  &  Pain& Own&-&87.00&85.18&88.32&-\\
         
         &   \multicolumn{1}{l}{\cite{10011731} } &SwinT &  Visual  &  Pain &UNBC-McMaster&95.25&-&-&-&- \\
        &   \multicolumn{1}{l}{\cite{zhao2022unaligned} } &Cross-modal Attention-GAN  &  Text,Audio,Visual  &  Depression&AVEC2019&-&-&-&-&3.56 \\
        &   \multicolumn{1}{l}{\cite{han2023spatial}}  &Self-supervised Learning  & Audio  & Depression&AVEC2017&80.00&76.00&65.00&92.00&-\\
          &    \multicolumn{1}{l}{\cite{liu2023net} }  &Self-attention Mechanism & Visual  & Depression &AVEC2014&-&-&-&-&6.04\\       
        &    \multicolumn{1}{l}{\cite{lu2023video} }&TS-ConvNet-CSA   &  Visual  &  Pain &DFEPN &66.20&-&-&-&-\\
        &  \multicolumn{1}{l}{\cite{wang2023multimodal} }&ALBERT-AGG   &  Text,Visual  &  Depression &Chinese Sina Weibo&93.82&91.00&92.52&88.58&-\\      
        &  \multicolumn{1}{l}{\cite{Anekar2023ExploringEA} } &CNN-NLTK  &  Audio, Visual  &  Depression &FER2013&77.00&-&80.00&91.00&-\\      
         &  \multicolumn{1}{l}{\cite{10271158} }  &LSTM-CNN-Attention Mechanism &  Text, Audio  &  Depression &E-DAIC&83.00&-&89.00&86.00&-\\
        \hline

        \multirow{15}*{P}
        
        &   \multicolumn{1}{l} {\cite{BAJESTANI2019277} }  &KNN &  EEG  &  Autism &Own&81.67&-&-&-&-\\
        &   \multicolumn{1}{l} {\cite{HADOUSH2019240} } &ANN  &  EEG  &  Autism &Own&97.20&-&-&-&-\\
         &   \multicolumn{1}{l} {\cite{8653893} }  &MDAE-SVM&  EEG,EM  & Depression &Own&83.42&-&-&-&-\\
         &   \multicolumn{1}{l} {\cite{ijerph17030971} }&LSDA-PNN    &  EEG  &  Autism &Own&98.70&-&-&-&-\\
         &   \multicolumn{1}{l} {\cite{BAYGIN2021104548} } &Lightweight deep networks-SVM &  EEG  &  Autism &Own&96.44&-&-&-&-\\
         
          &   \multicolumn{1}{l} {\cite{WADHERA2021102556} }  &VG-SVM  &  EEG  &  Autism &Own&94.19&-&-&-&-\\
        &   \multicolumn{1}{l} {\cite{ABDOLZADEGAN2020482} }&SVM  & EEG  &  Autism &Own&90.57&-&-&-&-\\
        &   \multicolumn{1}{l} {\cite{han2022multimodal} } &SDAE  &  EEG,ET  & Autism  &Own&95.56&-&-&-&-\\
        &    \multicolumn{1}{l}{\cite{xia2023end}}&MHSA-CNN   & EEG  & Depression &HUSM&91.06&-&-&-&- \\
        &   \multicolumn{1}{l}{\cite{torres2023evaluation}} &CNN-ROAR   &  EEG  &  Autism &Own&93.40&93.40&93.50&93.30&-\\
        &   \multicolumn{1}{l}{\cite{phan2023pain} } &CNN-BiLSTM  &  ECG,EDA  &  Pain &BioVid heat pain&84.80&-&-&-&-\\
        &   \multicolumn{1}{l}{\cite{10340623} } &ASGC   &  EEG  &  Depression &Own&84.27&84.22&-&-&-\\
        &   \multicolumn{1}{l}{\cite{10097883} }  &FL  &  EEG  &  Depression &Own&75.00&-&-&-&-\\
         &   \multicolumn{1}{l}{\cite{Zhang2023DepressionRB} }   &SVM &  ECG  &  Depression &Own&70.00&61.54&85.71&-&-\\
         \hline
         
      \multirow{4}*{B-P}        
       &   \multicolumn{1}{l} {\cite{chen2019attention}} &ResNet-LSTM  &  ET,Visual  &  Autism &Own&84.00&-&-&-&-\\          
          &   \multicolumn{1}{l} {\cite{ABDULHAMID20231965}}  &LSTM  &  EEG,Visual  &  Depression &Own&96.80&-&-&-&-\\         
         &  \multicolumn{1}{l}{\cite{qayyum2023high} }  &ViT & EEG,Audio  &Depression &MODMA&97.31&97.34&97.71&97.34 &-\\
         &  \multicolumn{1}{l}{\cite{10290732} } &Multimodal fusion-SVM  & EEG,Audio  &Depression &Own&86.78&-&-&-&- \\
        \bottomrule[2pt]
    \end{tabular}}
    \caption{A technical overview and performance comparison of affective computing approaches in healthcare over the last five years. B, P and B-P represent behaviour, physiological and behavioural-physiological, respectively. Own represents an unpublished dataset constructed by authors. Due to space constraints, we only list representative methods.}
    \label{tab: plain1}
\end{table*}

\section{Applications on Healthcare}
\begin{figure}[t]
    \centering
    \includegraphics[width=0.42\textwidth]{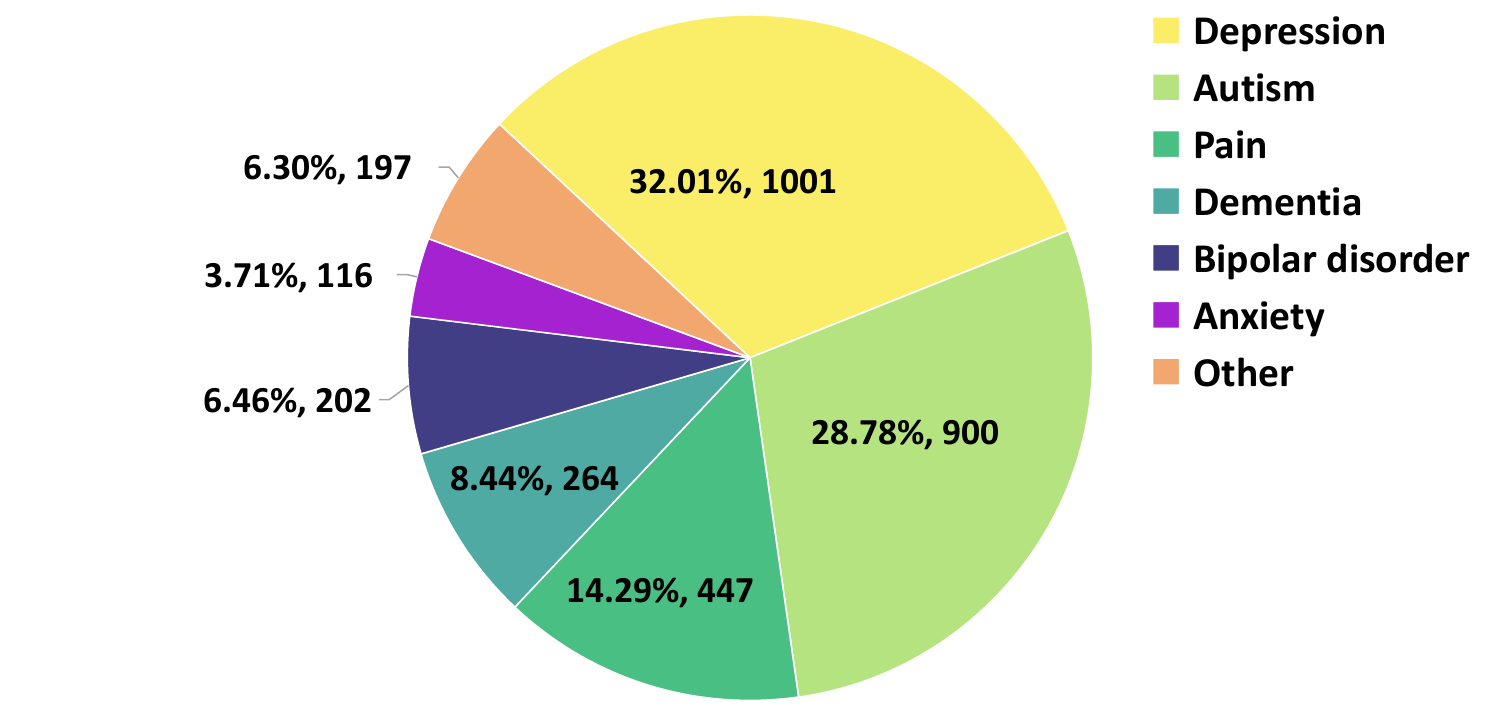}
    \caption{The statistic distribution of publications on various healthcare applications of affective computing from 2019 to 2023 based on Keyword search on web of science. We emphasize the top three most frequently explored applications.}
    \label{Fig.main3}
\end{figure}

In this section, the affective approaches for \emph{healthcare applications} are presented. 
Affective computing holds significant potential in various healthcare applications, offering new techniques for diagnosis, therapy, and treatment of emotion-related diseases by identifying patients' emotions.
We collect the studies of affective computing in healthcare applications that has been proposed in the last five years, and Figure \ref{Fig.main3} provides the statistics. 
In general, the application of affective computing technology in the field of healthcare involves depression diagnosis, autism recognition, pain level recognition, elderly dementia monitoring, stress monitoring, and intelligent medical systems.
For simplicity,
we review the top three most popular healthcare applications related to brain disorders with representative methods, including depression diagnosis, autism recognition, and pain level recognition. These applications can cover the major populations of current brain disorders, including children, adults, and the elderly. After the review of 3 popular applications, we also provide a brief review of the rest of the applications.

\subsection{Depression Diagnosis}




Depression, a widespread mental disorder, results from a complex interplay of social, psychological, and biological factors, leading to prolonged periods of low mood or reduced interest.
Traditional self-report-based diagnosis is subjective and prone to inaccuracies, often resulting in delays in treatment.  To tackle this, affective computing technology is applied for depression diagnosis, improving diagnosis objectivity and accuracy. Currently, affective computing-based depression diagnosis involves two categories: classification-based and regression-based depression diagnosis. 


Affective classification-based depression diagnosis aids healthcare workers and doctors in distinguishing between individuals with and without depression.
\cite{alsharif2022depression} used mel-frequency cepstral coefficients to extract patients' audio features and CNN to build a classification model to detect depression in Arabic audio data.
~\cite{cai2020feature} employed KNN as a depression classification detection model to distinguish depression patients from normal people by integrating different EEG data obtained under neutral, negative, and positive audio stimulation.~\cite{wang2023multimodal} proposed a multi-modal depression detection model with emotional knowledge graph, which integrated text, facial expressions, and other behaviors (\textit{e.g.}, the number of user posts, blog length, \textit{etc.}) to address the depression detection task.
\cite{pan2023integrating} used Transformer to classify depression and non-depression from audio signals and facial expressions.


Since regression allows access to continuous affective states, affective regression-based depression diagnosis has been proposed to predict the degree of depression or the severity of depressive symptoms in patients. \cite{zhou2020visually} proposed a deep regression network called DepressNet with facial depression data to predict depression degree. This significantly improves the latest performance of visual-based complex depression recognition.
~\cite{zhao2022unaligned} employed a cross-modal affective regression model to facilitate the learning of more accurate multimodal representations from text, audio, and facial expression videos for automatic depression severity assessment.

\textbf{Clinical Application}
With significant advancements in affective computing technology, depression diagnosis is now being employed in clinical monitoring. To prevent the recurrence of patients with depression, ~\cite{yin2022intelligent} designed an intelligent monitoring system based on a hybrid affective computing model (namely CNN-LSTM), aiming to provide recurrence monitoring for patients with depression within their home and daily environments. Additionally, it can be applied to assess new patients. The system includes user input, depression testing, intelligent monitoring, and connectivity to external wearables like dedicated voice acquisition devices and EEG devices. It also supports communication with online doctors and integration with external systems.

\textbf{Limitation}
Overall, depression diagnosis with affective computing is evolving from rough classification to continuous regression analyses, for dynamic clinical monitoring. However, it still faces challenges including subjectivity and heterogeneity of depression, longitudinal monitoring, and comprehensive assessment integrating genetic, psychosocial, and environmental factors. In addition, current affective computing technologies rely mainly on behavioural and physiological indicators while lacking clear biological indicators to support an objective diagnosis of depression, affecting diagnostic accuracy.

\subsection{Autism Recognition and Intervention}

Autism, a neurodevelopmental disorder, is typically characterized by emotional and social difficulties. Early diagnosis is crucial for facilitating intervention and treatment. Affective computing can aid experts in promptly assessing the emotional state of individuals with autism during interactions, enhancing diagnostic accuracy and positive emotional interventions. As a result, affective computing-based autism diagnosis and intervention are gaining increasing attention.

Currently, autism diagnosis primarily employs emotion classification models to identify Autism spectrum disorder (ASD).~\cite{negin2021vision} proposed a non-invasive visual assistance method with human action classification to facilitate the diagnosis of autism.~\cite{rahman2022early} employed the MobileNet to detect childhood autism through their facial expressions in a transfer learning manner.
~\cite{wei2023vision} proposed a lightweight, conventional classification model to recognize autism-related behaviors in facial videos.
~\cite{han2022multimodal} build a stacked denoising autoencoder to identify ASD in children from EEG and ET fusion data.~\cite{li2021two} combined audio and facial expression images to diagnose ASD in children, and used ResNet50 and ResNet18 as classifiers for a two-stage emotion classification, thus improving the accuracy of autism diagnosis

\textbf{Clinical Application}
In clinical applications of autism intervention, leveraging affective computing technology enhances virtual reality (VR) for human-computer interaction (HCI) scenarios addressing challenges caused by a shortage of autism treatment professionals.~\cite{manjuincreasing} employ VR-assisted system combined with wearable multi-model sensing technologies, to collect physiological signals and game performance data during HCI training. Then, it employs a machine learning model to identify ASD children, assessing the diagnosis, severity, social behavioral intervention, and treatment of ASD with multiple assessment scales. 

\textbf{Limitation}
Despite some progress in autism diagnosis and intervention, current approaches are still in their infancy and face challenges such as data privacy concerns in children, difficulties in data collection, assessment complexity, and emotion model bias in autism. In addition,  Autism may have other co-existing mental health disorders, such as attention deficit hyperactivity disorder (ADHD), complicating affective computing models for accurate diagnosis.  



\subsection{Pain Level Recognition}
Pain is the body's intricate physical and psychological emotional response to underlying injury or illness. Accurate identification of pain is crucial in medicine, enabling medicine professionals to formulate effective treatment plans for enhancing a patient's quality of life. Research on pain level recognition through affective computing has been a prominent and challenging issue. The field of pain level recognition employs affective computing techniques for classification and regression tasks, aimed at diagnosing pain patients and obtaining their pain intensity, respectively. 

Affective classification-based pain level recognition can assist medical professionals in accurately determining the patient's pain location, thereby significantly optimizing consultation time. For instance, ~\cite{vallez2022adults} identified joint pain from facial expression images with the help of a pre-trained CNN classification model.~\cite{chen2022scalp} used multi-layer CNN classifies the EEG singals in resting and pain states during daily activities.~\cite{lu2023video} identified newborns' pain levels based on their facial expression videos with the help of Softmax. Similar to~\cite{lu2023video}, \cite{phan2023pain} employed Softmax to recognition pain levels from EDA and ECG signals. 

To obtain a continuous numerical output for the pain degree of patients, pain level recognition based on affective regression is developed. \cite{thiam2020multimodal} employed feed-forward neural networks as a regression model for discerning pain level intensity based on emotion-related physiological signals (\textit{i.e.}, EDA, EMG, and ECG).
Besides, \cite{jiang2024personalized} used a non-linear neural network with Sigmoid for both classification and regression tasks, distinguishing between pain and non-pain in patients and detecting pain intensity based on ECG, EDA, and ECG.
\textbf{Clinical Application}
In a clinical setting, pain analysis is critical to a patient's recovery. \cite{ghosh2023smart} proposed an emotion analysis system based on deep learning and statistical learning, 
to analyze facial expressions images of patients for detecting pain levels of patients. 
In addition, the system has the ability to perform pain detection and recognition on resource-constrained devices, which provides a strong support for the intelligent healthcare field.

\textbf{Limitation}
Pain level recognition has developed from initial classification of pain versus non-pain to the recognition of pain intensity, from clinical diagnosis to real-time monitoring. 
However, pain recognition and assessment still faces significant challenges such as individual differences, emotional differences, standard calibration, and so on. In addition, some pain symptoms may lack distinct behavioural and physiological features, further complicating identification.


\subsection{Other Related Healthcare Applications}
Affective computing has also been applied to other clinical applications such as Bipolar disorder~\cite{baki2022multimodal}, elderly companionship and monitoring~\cite{meng2021hybrid}, smart medicine~\cite{ayata2020emotion} etc. For convenience, we combine some of them into one section for overview.

\underline{Bipolar disorder} is a mental health disorder that causes mood swings ranging from depression to mania.~\cite{baki2022multimodal} created a multimodal decision system for three level mania classification based on recordings of patients' audio, text, and facial expression videos. \underline{Elderly monitoring and dementia diagnosis} are crucial in an aging population, especially during illness. To accurately monitor the emotion state of the elderly,~\cite{meng2021hybrid} introduced an emotion-aware medicine monitoring system based on brain waves. 
\underline{Intelligent medicine systems} offer an alternative to doctor shortages. \cite{ayata2020emotion} proposed an emotion recognition-based intelligent medicine system for emotional care by collecting and analyzing multiple physiological signals from patients.




\section{Challenges and Opportunities}
Despite breakthroughs, several challenges remain, yet there are also related opportunities for future development. 


\subsection{Patient Data Privacy and Ethics} 
\paragraph{Challenge} Data privacy issues have been well-known in the big data era, and it is particularly important for the healthcare sector. This might be attributed to the fact that patients have to share extremely sensitive information about their own bodies. As a result, the privacy of patient clinical data in affective computing is a crucial ethical concern, especially for children's information. Ensuring confidentiality involves implementing robust measures in data transmission, storage, and usage. The challenge is to retain critical emotion-related information while adhering to ethical and moral regulations. Therefore, a proper balance needs to be found to ensure privacy and the effectiveness of data analysis.

\paragraph{Opportunities}  
Some potential future opportunities may include the exploration of advanced privacy-protecting techniques such as federated learning \cite{rieke2020future} and secure multi-party computing \cite{liu2020privacy}. Federated learning allows for model training without sharing raw data, and aggregating models without exposing individual data. Secure multi-party computing allows calculations to be performed between multiple parties while maintaining the privacy of the data. These technologies can ensure that patients' clinical data is adequately protected.


\subsection{Emotion Bias and Fairness in Clinical Data } 

\paragraph{Challenge} 
In clinical and other healthcare settings, the collection and annotation of sentiment data exhibit a natural bias due to population sentiment expression heterogeneity and annotator subjectivity. Different from data bias issues in common large datasets, the data from healthcare suffers from low data volume and larger diversities of biased factors. For example, environment, age, occupation, and race can all affect the expression and the labelling of emotions \cite{liu2022mafw}. 
Furthermore, preferences and concerns during data collection and labeling also vary significantly among doctors. These factors could lead to high emotion biases and unfairness in training AI models. As a result, it is needed to address and reduce the bias for a fair and unbiased understanding of emotions across diverse populations.  

\paragraph{Opportunities} 
To lower the biases in affective computing under limited labelled databases, some research is exploring unsupervised/self-supervised learning algorithms to reduce reliance on affective labels. These algorithms can learn emotional representations from unlabeled data, reducing the need for large-scale labeled datasets \cite{han2023spatial,han2022multimodal}. Furthermore, the introduction of domain adaptive learning techniques can improve the generality of affective computing models and mitigate affective biases between different cultures and demographic groups. 

\subsection{Fine-grained Health-related Emotions}  
\paragraph{Challenge} Most existing healthcare applications rely on single, simplistic affective models like six-class or three-class emotion classification models \cite{ameer2023multi}. These models fall short of simulating rich emotions from real patients who may be undergoing complicated treatments, making it challenging for doctors to make accurate judgments. As a result, developing fine-grained health-related emotion models for clinical applications remains a key unresolved issue.

\paragraph{Opportunities}  
Recently, several researchers have proposed composite face expression models based on linguistic descriptions \cite{liu2022mafw}. We believe that this approach facilitates the description of changes in emotional details, thereby guiding doctors to make more informed medical diagnoses. Consequently, constructing multimodal fine-grained emotion models in healthcare applications emerges as a future development direction.

\subsection{Real-time Diagnosis with Affective Computing} 
\paragraph{Challenge} Some medicine applications demand real-time emotional analysis and diagnosis, involving the rapid processing and analysis of large data sets while ensuring accurate emotion recognition, such as mental health monitoring or emergency response systems. Developing efficient algorithms and infrastructure for real-time processing without compromising accuracy is a significant challenge.

 


\paragraph{Opportunities}  Exploring adaptive algorithms and edge computing systems facilitates the capability of real-time emotional analysis with minimal latency. Adaptive algorithms can be directly adjusted and optimized according to different needs to improve the efficiency and accuracy of sentiment analysis. Edge computing systems can discretize computing tasks to edge devices, reducing latency in data transmission and processing for faster real-time sentiment analysis.

\subsection{Large Foundation Model-related Applications} 
\paragraph{Challenge}
With the development of visual-linguistic large foundation models, such as GPT-4 \cite{rathje2023gpt}, the significance of large foundation models has been demonstrated across various application domains. Consequently, constructing an affective large foundation model specifically for healthcare could be beneficial to enhance a wider range of clinical applications.
However, the acquisition and annotation of specified large foundation models for healthcare and clinical data remain challenging, posing a significant hurdle in the development of  healthcare 
affective foundation model.

\paragraph{Opportunities}  
Leveraging existing visual-language foundation models to construct the affective large-scale model through transfer learning and cross-modal prompt learning could reduce the dependence on large amounts of training data, thus enhancing the reusability of these models for diverse application tasks \cite{liu2024emollms}. 
This method can not only improve the effect and generalization ability of affective models but also provide a common basis for emotion recognition in different domains and tasks. 

\section{Conclusion}
This paper provides a comprehensive survey of the application of affective computing in the field of healthcare.
Specifically, we provide an overview of the developments in affective computing for healthcare, covering behavior data-based, psychological data-based, and behavior-psychological data-based datasets and approaches. Next, we introduce key healthcare applications, highlighting the top three most frequently used, as well as other related applications. Finally, we summarize the most potential challenges and opportunities in the development of affective computing in healthcare.
We believe that this review helps to provide academic and industrial researchers with a comprehensive understanding of the latest advances in affective computing-based healthcare and provides them with guidance.

\begin{spacing}{0.80}
  \bibliographystyle{named} 
  \small
  \bibliography{ijcai24}

\begin{thebibliography}{}

\bibitem[\protect\citeauthoryear{Abdolzadegan \bgroup \em et al.\egroup }{2020}]{ABDOLZADEGAN2020482}
D.~Abdolzadegan, MH. Moattar, and M.~Ghoshuni.
\newblock A robust method for early diagnosis of autism spectrum disorder from eeg signals based on feature selection and dbscan method.
\newblock {\em Biocybern Biomed Eng}, 2020.

\bibitem[\protect\citeauthoryear{Alelaiwi}{2019}]{alelaiwi2019multimodal}
A.~Alelaiwi.
\newblock Multimodal patient satisfaction recognition for smart healthcare.
\newblock {\em IEEE Access}, 2019.

\bibitem[\protect\citeauthoryear{Alghifari \bgroup \em et al.\egroup }{2023}]{alghifari2023development}
MF. Alghifari, TS. Gunawan, and M.~Kartiwi.
\newblock Development of sorrow analysis dataset for speech depression prediction.
\newblock In {\em I2MTC}, 2023.

\bibitem[\protect\citeauthoryear{Alizadeh and Tabibian}{2021}]{Alizadeh2021APS}
M.~Alizadeh and S.~Tabibian.
\newblock A persian speaker-independent dataset to diagnose autism infected children based on speech processing techniques.
\newblock {\em ICSPIS}, 2021.

\bibitem[\protect\citeauthoryear{Alsharif \bgroup \em et al.\egroup }{2022}]{alsharif2022depression}
Z.~Alsharif, S.~Elhag, and S.~Alfakeh.
\newblock Depression detection in arabic using speech language recognition.
\newblock In {\em CDMA}, 2022.

\bibitem[\protect\citeauthoryear{Ameer \bgroup \em et al.\egroup }{2023}]{ameer2023multi}
I.~Ameer, N.~Bölücü, MHF. Siddiqui, B.~Can, et~al.
\newblock Multi-label emotion classification in texts using transfer learning.
\newblock {\em Expert Syst. Appl}, 2023.

\bibitem[\protect\citeauthoryear{Anekar \bgroup \em et al.\egroup }{2023}]{Anekar2023ExploringEA}
D.~Anekar, Y.~Deshpande, R.~Suryawanshi, R.~Waman, V.~Divekar, and R.~Salunke.
\newblock Exploring emotion and sentiment landscape of depression: A multimodal analysis approach.
\newblock {\em GCAT}, 2023.

\bibitem[\protect\citeauthoryear{Ayata \bgroup \em et al.\egroup }{2020}]{ayata2020emotion}
D.~Ayata, Y.~Yaslan, and ME. Kamasak.
\newblock Emotion recognition from multimodal physiological signals for emotion aware healthcare systems.
\newblock {\em J Med Biol Eng}, 2020.

\bibitem[\protect\citeauthoryear{Bajestani \bgroup \em et al.\egroup }{2019}]{BAJESTANI2019277}
GS. Bajestani, M.~Behrooz, AG. Khani, M.~Nouri-Baygi, and A.~Mollaei.
\newblock Diagnosis of autism spectrum disorder based on complex network features.
\newblock {\em Comput Methods Programs Biomed}, 2019.

\bibitem[\protect\citeauthoryear{Baki \bgroup \em et al.\egroup }{2022}]{baki2022multimodal}
P.~Baki, H.~Kaya, E~.Çiftçi, and H.~Güleç.
\newblock A multimodal approach for mania level prediction in bipolar disorder.
\newblock {\em IEEE Transactions on Affective Computing}, 2022.

\bibitem[\protect\citeauthoryear{Baygin \bgroup \em et al.\egroup }{2021}]{BAYGIN2021104548}
M.~Baygin, S.~Dogan, T.~Tuncer, PD. Barua, O.~Faust, N.~Arunkumar, EW. Abdulhay, EE~Palmer, and UR. Acharya.
\newblock Automated asd detection using hybrid deep lightweight features extracted from eeg signals.
\newblock {\em Comput. Biol. Med}, 2021.

\bibitem[\protect\citeauthoryear{Cai \bgroup \em et al.\egroup }{2020a}]{cai2020feature}
H.~Cai, Z.~Qu, Z.~Li, Y.~Zhang, X.~Hu, and B.~Hu.
\newblock Feature-level fusion approaches based on multimodal eeg data for depression recognition.
\newblock {\em Inf Fusion}, 2020.

\bibitem[\protect\citeauthoryear{Cai \bgroup \em et al.\egroup }{2020b}]{Cai2020AMO}
H.~Cai, Z.~Yuan, Y.~Gao, S.~Sun, N.~Li, F.~Tian, H.~Xiao, J.~Li, Z.~Yang, X.~Li, Q.~Zhao, Z.~Liu, Z.~Yao, et~al.
\newblock A multi-modal open dataset for mental-disorder analysis.
\newblock {\em Sci. Data}, 2020.

\bibitem[\protect\citeauthoryear{Calabrò \bgroup \em et al.\egroup }{2021}]{calabro2021m}
G.~Calabrò, A.~Bizzego, S.~Cainelli, C.~Furlanello, and P.~Venuti.
\newblock M-ms: A multi-modal synchrony dataset to explore dyadic interaction in asd.
\newblock {\em Progresses in Artificial Intelligence and Neural Systems}, 2021.

\bibitem[\protect\citeauthoryear{Chen and Luo}{2023}]{2}
X.~Chen and T.~Luo.
\newblock Catching elusive depression via facial micro-expression recognition.
\newblock {\em IEEE Commun. Mag.}, 2023.

\bibitem[\protect\citeauthoryear{Chen and Zhao}{2019}]{chen2019attention}
S.~Chen and Q.~Zhao.
\newblock Attention-based autism spectrum disorder screening with privileged modality.
\newblock In {\em ICCV}, 2019.

\bibitem[\protect\citeauthoryear{Chen \bgroup \em et al.\egroup }{2022}]{chen2022scalp}
D.~Chen, H.~Zhang, PT. Kavitha, FL. Loy, SH. Ng, C.~Wang, KS. Phua, SY. Tjan, SY. Yang, and C.~Guan.
\newblock Scalp eeg-based pain detection using convolutional neural network.
\newblock {\em IEEE Trans. Neural Syst. Rehab. Eng.}, 2022.

\bibitem[\protect\citeauthoryear{Dessai and Usgaonkar}{2022}]{dessai2022depression}
S.~Dessai and SS. Usgaonkar.
\newblock Depression detection on social media using text mining.
\newblock In {\em INCET}, 2022.

\bibitem[\protect\citeauthoryear{Duan \bgroup \em et al.\egroup }{2019}]{emcasd}
H.~Duan, G.~Zhai, X.~Min, Z.~Che, Y.~Fang, X.~Yang, J.~Gutiérrez, and PL. Callet.
\newblock A dataset of eye movements for the children with autism spectrum disorder.
\newblock In {\em ACM MM}, 2019.

\bibitem[\protect\citeauthoryear{Fernandes-Magalhaes \bgroup \em et al.\egroup }{2023}]{fernandes2023pain}
R.~Fernandes-Magalhaes, A.~Carpio, D.~Ferrera, D.~Van Ryckeghem, I.~Peláez, P.~Barjola, et~al.
\newblock Pain emotion faces database (pemf): Pain-related micro-clips for emotion research.
\newblock {\em Behav Res Methods}, 2023.

\bibitem[\protect\citeauthoryear{Ghosh \bgroup \em et al.\egroup }{2023}]{ghosh2023smart}
A.~Ghosh, S.~Umer, MK. Khan, RK. Rout, and BC. Dhara.
\newblock Smart sentiment analysis system for pain detection using cutting edge techniques in a smart healthcare framework.
\newblock {\em Cluster Computing}, 2023.

\bibitem[\protect\citeauthoryear{Giuntini \bgroup \em et al.\egroup }{2020}]{giuntini2020review}
FT. Giuntini, MT. Cazzolato, MJD. dos Reis, et~al.
\newblock A review on recognizing depression in social networks: challenges and opportunities.
\newblock {\em Journal of Ambient Intelligence and Humanized Computing}, 2020.

\bibitem[\protect\citeauthoryear{Hadoush \bgroup \em et al.\egroup }{2019}]{HADOUSH2019240}
H.~Hadoush, M.~Alafeef, and E.~Abdulhay.
\newblock Eeg analysis using empirical mode decomposition and second order difference plot.
\newblock {\em Behavioural Brain Research}, 2019.

\bibitem[\protect\citeauthoryear{Hamid \bgroup \em et al.\egroup }{2023}]{ABDULHAMID20231965}
DSBA. Hamid, SB. Goyal, and P.~Bedi.
\newblock Integration of deep learning for improved diagnosis of depression using eeg and facial features.
\newblock {\em Mater. Today.}, 2023.

\bibitem[\protect\citeauthoryear{Han \bgroup \em et al.\egroup }{2022}]{han2022multimodal}
J.~Han, G.~Jiang, G.~Ouyang, and X.~Li.
\newblock A multimodal approach for identifying autism spectrum disorders in children.
\newblock {\em IEEE Trans. Neural Syst. Rehab. Eng.}, 2022.

\bibitem[\protect\citeauthoryear{Han \bgroup \em et al.\egroup }{2023}]{han2023spatial}
Z.~Han, Y.~Shang, Z.~Shao, J.~Liu, G.~Guo, T.~Liu, H.~Ding, and Q.~Hu.
\newblock Spatial-temporal feature network for speech-based depression recognition.
\newblock {\em IEEE Trans Cogn Dev Syst}, 2023.

\bibitem[\protect\citeauthoryear{Hossain and Muhammad}{2019}]{hossain2019emotion}
MS. Hossain and G.~Muhammad.
\newblock Emotion recognition using secure edge and cloud computing.
\newblock {\em Information Sciences}, 2019.

\bibitem[\protect\citeauthoryear{Ji \bgroup \em et al.\egroup }{2023}]{ji2023towards}
Z.~Ji, T.~Yu, Y.~Xu, N.~Lee, E.~Ishii, and P.~Fung.
\newblock Towards mitigating llm hallucination via self reflection.
\newblock In {\em EMNLP}, 2023.

\bibitem[\protect\citeauthoryear{Jiang \bgroup \em et al.\egroup }{2022}]{jiang2022mmda}
Y.~Jiang, Z.~Zhang, and X.~Sun.
\newblock Mmda: A multimodal dataset for depression and anxiety detection.
\newblock In {\em ICPR}, 2022.

\bibitem[\protect\citeauthoryear{Jiang \bgroup \em et al.\egroup }{2024}]{jiang2024personalized}
M.~Jiang, R.~Rosio, S.~Salanterä, AM. Rahmani, and other.
\newblock Personalized and adaptive neural networks for pain detection from multi-modal physiological features.
\newblock {\em Expert Syst. Appl}, 2024.

\bibitem[\protect\citeauthoryear{Lan \bgroup \em et al.\egroup }{2023}]{10340623}
YT. Lan, D.~Peng, W.~Liu, Y.~Luo, Z.~Mao, WL. Zheng, and BL. Lu.
\newblock Investigating emotion eeg patterns for depression detection with attentive simple graph convolutional network.
\newblock In {\em EMBC}, 2023.

\bibitem[\protect\citeauthoryear{Li \bgroup \em et al.\egroup }{2021}]{li2021two}
J.~Li, A.~Bhat, and R.~Barmaki.
\newblock A two-stage multi-modal affect analysis framework for children with autism spectrum disorder.
\newblock {\em arXiv preprint arXiv:2106.09199}, 2021.

\bibitem[\protect\citeauthoryear{Liu \bgroup \em et al.\egroup }{2020}]{liu2020privacy}
J.~Liu, Y.~Tian, Y.~Zhou, Y.~Xiao, and N.~Ansari.
\newblock Privacy preserving distributed data mining based on secure multi-party computation.
\newblock {\em Computer Communications}, 2020.

\bibitem[\protect\citeauthoryear{Liu \bgroup \em et al.\egroup }{2022}]{liu2022mafw}
Y.~Liu, W.~Dai, C.~Feng, W.~Wang, G.~Yin, J.~Zeng, and S.~Shan.
\newblock Mafw: A large-scale and multi-modal and compound affective database for dynamic facial expression recognition in the wild.
\newblock In {\em Proc. 30th ACM Int. Conf. Multimedia}, 2022.

\bibitem[\protect\citeauthoryear{Liu \bgroup \em et al.\egroup }{2023}]{liu2023net}
Z.~Liu, X.~Yuan, Y.~Li, Z.~Shangguan, L.~Zhou, et~al.
\newblock Pra-net: Part-and-relation attention network for depression recognition from facial expression.
\newblock {\em Comput. Biol. Med}, 2023.

\bibitem[\protect\citeauthoryear{Liu \bgroup \em et al.\egroup }{2024}]{liu2024emollms}
Z.~Liu, K.~Yang, T.~Zhang, Q.~Xie, Z.~Yu, and S.~Ananiadou.
\newblock Emollms: A series of emotional large language models and annotation tools for comprehensive affective analysis.
\newblock {\em arXiv preprint arXiv:2401.08508}, 2024.

\bibitem[\protect\citeauthoryear{Lu \bgroup \em et al.\egroup }{2023}]{lu2023video}
G.~Lu, H.~Chen, J.~Wei, X.~Li, X.~Zheng, H.~Leng, Y.~Lou, and J.~Yan.
\newblock Video-based neonatal pain expression recognition with cross-stream attention.
\newblock {\em Multimed. Tools. Appl}, 2023.

\bibitem[\protect\citeauthoryear{Mallol-Ragolta \bgroup \em et al.\egroup }{2020}]{9320205}
A.~Mallol-Ragolta, S.~Liu, N.~Cummins, and B.~Schuller.
\newblock A curriculum learning approach for pain intensity recognition from facial expressions.
\newblock In {\em FG}, 2020.

\bibitem[\protect\citeauthoryear{Manju \bgroup \em et al.\egroup }{2023}]{manjuincreasing}
T.~Manju, Magesh, S.~Padmavathi, and Durairaj.
\newblock Increasing the social interaction of autism child using virtual reality intervention (vri).
\newblock {\em TALLIP}, 2023.

\bibitem[\protect\citeauthoryear{Melinda \bgroup \em et al.\egroup }{2023}]{melinda2023novel}
M.~Melinda, A.~Ahmadiar, M.~Oktiana, M.~ShadiqAdiNugraha, MAL. Qadrillah, and Y.~Yunidar.
\newblock A novel autism spectrum disorder children dataset based on thermal imaging.
\newblock In {\em ICCCE}, 2023.

\bibitem[\protect\citeauthoryear{Meng \bgroup \em et al.\egroup }{2021}]{meng2021hybrid}
W.~Meng, Y.~Cai, LT. Yang, and WY. Chiu.
\newblock Hybrid emotion-aware monitoring system based on brainwaves for internet of medical things.
\newblock {\em IEEE Internet Things J}, 2021.

\bibitem[\protect\citeauthoryear{Mohammadi \bgroup \em et al.\egroup }{2019}]{8786540}
Y.~Mohammadi, M.~Hajian, and MH. Moradi.
\newblock Discrimination of depression levels using machine learning methods on eeg signals.
\newblock In {\em ICEE}, 2019.

\bibitem[\protect\citeauthoryear{Mohan \bgroup \em et al.\egroup }{2022}]{mohan2022edge}
HM. Mohan, HCS. Kumara, SH. Mallikarjun, and AY. Prasad.
\newblock Edge artificial intelligence-based facial pain recognition during myocardial infarction.
\newblock {\em JAMRIS}, 2022.

\bibitem[\protect\citeauthoryear{Negin \bgroup \em et al.\egroup }{2021}]{negin2021vision}
F.~Negin, B.~Ozyer, S.~Agahian, S.~Kacdioglu, and GT. Ozyer.
\newblock Vision-assisted recognition of stereotype behaviors for early diagnosis of autism spectrum disorders.
\newblock {\em Neurocomputing}, 2021.

\bibitem[\protect\citeauthoryear{Othmani \bgroup \em et al.\egroup }{2021}]{othmani2021towards}
A.~Othmani, D.~Kadoch, K.~Bentounes, E.~Rejaibi, R.~Alfred, and A.~Hadid.
\newblock Towards robust deep neural networks for affect and depression recognition from speech.
\newblock In {\em Pattern Recognition. ICPR International Workshops and Challenges: Virtual Event and January 10--15 and 2021 and Proceedings and Part II}, 2021.

\bibitem[\protect\citeauthoryear{Pan \bgroup \em et al.\egroup }{2023}]{pan2023integrating}
Y.~Pan, Y.~Shang, Z.~Shao, T.~Liu, G.~Guo, and H.~Ding.
\newblock Integrating deep facial priors into landmarks for privacy preserving multimodal depression recognition.
\newblock {\em IEEE Trans Affect Comput.}, 2023.

\bibitem[\protect\citeauthoryear{Peng \bgroup \em et al.\egroup }{2020}]{peng2020pain}
X.~Peng, D.~Huang, and H.~Zhang.
\newblock Pain intensity recognition via multi-scale deep network.
\newblock {\em IET Image Processing}, 2020.

\bibitem[\protect\citeauthoryear{Pham \bgroup \em et al.\egroup }{2020}]{ijerph17030971}
TH. Pham, J.~Vicnesh, JKE. Wei, SL. Oh, N.~Arunkumar, EW. Abdulhay, EJ. Ciaccio, and UR. Acharya.
\newblock Autism spectrum disorder diagnostic system using hos bispectrum with eeg signals.
\newblock {\em International Journal of Environmental Research and Public Health}, 2020.

\bibitem[\protect\citeauthoryear{Phan \bgroup \em et al.\egroup }{2023}]{phan2023pain}
KN. Phan, NK. Iyortsuun, S.~Pant, HJ. Yang, and SH. Kim.
\newblock Pain recognition with physiological signals using multi-level context information.
\newblock {\em IEEE Access}, 2023.

\bibitem[\protect\citeauthoryear{Qayyum \bgroup \em et al.\egroup }{2023}]{qayyum2023high}
A.~Qayyum, I.~Razzak, M.~Tanveer, M.~Mazher, and B.~Alhaqbani.
\newblock High-density electroencephalography and speech signal based deep framework for clinical depression diagnosis.
\newblock {\em IEEE ACM Trans Comput Bi}, 2023.

\bibitem[\protect\citeauthoryear{Rahman and Booma}{2022}]{rahman2022early}
LA. Rahman and PM. Booma.
\newblock The early detection of autism within children through facial recognition; a deep transfer learning approach.
\newblock In {\em NTIC}, 2022.

\bibitem[\protect\citeauthoryear{Rathje \bgroup \em et al.\egroup }{2023}]{rathje2023gpt}
S.~Rathje, DM. Mirea, I.~Sucholutsky, R.~Marjieh, et~al.
\newblock Gpt is an effective tool for multilingual psychological text analysis.
\newblock 2023.

\bibitem[\protect\citeauthoryear{Rejaibi \bgroup \em et al.\egroup }{2019a}]{rejaibi2019clinical}
E.~Rejaibi, D.~Kadoch, K.~Bentounes, R.~Alfred, M.~Daoudi, et~al.
\newblock Clinical depression and affect recognition with emoaudionet.
\newblock {\em arXiv preprint arXiv:1911.00310}, 2019.

\bibitem[\protect\citeauthoryear{Rejaibi \bgroup \em et al.\egroup }{2019b}]{Rejaibi2019ClinicalDA}
E.~Rejaibi, D.~Kadoch, K.~Bentounes, R.~Alfred, M.~Daoudi, et~al.
\newblock Clinical depression and affect recognition with emoaudionet.
\newblock {\em ArXiv}, 2019.

\bibitem[\protect\citeauthoryear{Rieke \bgroup \em et al.\egroup }{2020}]{rieke2020future}
N.~Rieke, J.~Hancox, W.~Li, F.~Milletari, HR. Roth, et~al.
\newblock The future of digital health with federated learning.
\newblock {\em NPJ digital medicine}, 2020.

\bibitem[\protect\citeauthoryear{Salekin \bgroup \em et al.\egroup }{2021}]{salekin2021multimodal}
MS. Salekin, G.~Zamzmi, J.~Hausmann, and D.~Goldgof….
\newblock Multimodal neonatal procedural and postoperative pain assessment dataset.
\newblock {\em Data in Brief}, 2021.

\bibitem[\protect\citeauthoryear{Shen \bgroup \em et al.\egroup }{2023}]{10097883}
J.~Shen, Y.~Zhang, H.~Liang, Z.~Zhao, K.~Zhu, K.~Qian, Q.~Dong, X.~Zhang, and B.~Hu.
\newblock Depression recognition from eeg signals using an adaptive channel fusion method via improved focal loss.
\newblock {\em IEEE J. Biomed. Health. Inf.}, 2023.

\bibitem[\protect\citeauthoryear{Smith \bgroup \em et al.\egroup }{2020}]{smith2020vocal}
M.~Smith, BJ. Dietrich, E.~Bai, and HJ. Bockholt.
\newblock Vocal pattern detection of depression among older adults.
\newblock {\em International journal of mental health nursing}, 2020.

\bibitem[\protect\citeauthoryear{Sousa \bgroup \em et al.\egroup }{2023}]{sousa2023introducing}
A.~Sousa, K.~Young, M.~D'aquin, M.~Zarrouk, and J.~Holloway.
\newblock Introducing calmed: Multimodal annotated dataset for emotion detection in children with autism.
\newblock In {\em International Conference on Human-Computer Interaction}, 2023.

\bibitem[\protect\citeauthoryear{Thiam \bgroup \em et al.\egroup }{2020}]{thiam2020multimodal}
P.~Thiam, HA. Kestler, and F.~Schwenker.
\newblock Multimodal deep denoising convolutional autoencoders for pain intensity classification based on physiological signals.
\newblock In {\em ICPRAM}, 2020.

\bibitem[\protect\citeauthoryear{Torres \bgroup \em et al.\egroup }{2023}]{torres2023evaluation}
JMM. Torres, S.~Medina-DeVilliers, T.~Clarkson, et~al.
\newblock Evaluation of interpretability for deep learning algorithms in eeg emotion recognition: A case study in autism.
\newblock {\em Artificial Intelligence in Medicine}, 2023.

\bibitem[\protect\citeauthoryear{Vallez \bgroup \em et al.\egroup }{2022}]{vallez2022adults}
N.~Vallez, J.~Ruiz-Santaquiteria, O.~Deniz, J.~Hughes, S.~Robertson, K.~Hoti, and G.~Bueno.
\newblock Adults’ pain recognition via facial expressions using cnn-based au detection.
\newblock In {\em ICIAP}, 2022.

\bibitem[\protect\citeauthoryear{Wadhera and Kakkarl}{2021}]{WADHERA2021102556}
T.~Wadhera and D.~Kakkarl.
\newblock Social cognition and functional brain network in autism spectrum disorder: Insights from eeg graph-theoretic measures.
\newblock {\em Biomedical Signal Processing and Control}, 2021.

\bibitem[\protect\citeauthoryear{Wang \bgroup \em et al.\egroup }{2021}]{9688837}
X.~Wang, S.~Zhao, and Y.~Wang.
\newblock Bimodal emotion recognition for the patients with depression.
\newblock In {\em ICSIP}, 2021.

\bibitem[\protect\citeauthoryear{Wang \bgroup \em et al.\egroup }{2023a}]{10290732}
X.~Wang, X.~Wan, Z.~Ning, Z.~Qie, J.~Li, and Y.~Xiao.
\newblock A multimodal fusion depression recognition assisted decision-making system based on eeg and speech signals.
\newblock In {\em CCCI}, 2023.

\bibitem[\protect\citeauthoryear{Wang \bgroup \em et al.\egroup }{2023b}]{wang2023multimodal}
Z.~Wang, B.~Deng, X.~Shu, and J.~Shu.
\newblock Multimodal depression detection model fusing emotion knowledge graph.
\newblock In {\em ICAIBD}, 2023.

\bibitem[\protect\citeauthoryear{Wei \bgroup \em et al.\egroup }{2023}]{wei2023vision}
P.~Wei, D.~Ahmedt-Aristizabal, H.~Gammulle, S.~Denman, and MA. Armin.
\newblock Vision-based activity recognition in children with autism-related behaviors.
\newblock {\em Heliyon}, 2023.

\bibitem[\protect\citeauthoryear{Xia \bgroup \em et al.\egroup }{2023}]{xia2023end}
M.~Xia, Y.~Zhang, Y.~Wu, and X.~Wang.
\newblock An end-to-end deep learning model for eeg-based major depressive disorder classification.
\newblock {\em IEEE Access}, 2023.

\bibitem[\protect\citeauthoryear{Xu \bgroup \em et al.\egroup }{2023}]{10271158}
X.~Xu, G.~Zhang, Q.~Lu, and X.~Mao.
\newblock Multimodal depression recognition that integrates audio and text.
\newblock In {\em ISCEIC}, 2023.

\bibitem[\protect\citeauthoryear{Yan \bgroup \em et al.\egroup }{2020}]{yan2020fenp}
J.~Yan, G.~Lu, X.~Li, W.~Zheng, C.~Huang, Z.~Cui, Y.~Zong, M.~Chen, Q.~Hao, Y.~Liu, J.~Zhu, and H.~Li.
\newblock Fenp: a database of neonatal facial expression for pain analysis.
\newblock {\em IEEE Trans Affect Comput.}, 2020.

\bibitem[\protect\citeauthoryear{Yao \bgroup \em et al.\egroup }{2022}]{yao2022d4}
B.~Yao, C.~Shi, L.~Zou, L.~Dai, M.~Wu, L.~Chen, Z.~Wang, and K.~Yu.
\newblock D4: a chinese dialogue dataset for depression-diagnosis-oriented chat.
\newblock {\em arXiv preprint arXiv:2205.11764}, 2022.

\bibitem[\protect\citeauthoryear{Yildirim-Celik \bgroup \em et al.\egroup }{2022}]{1}
H.~Yildirim-Celik, S.~Eroglu, and K.~Oguz...
\newblock Emotional context effect on recognition of varying facial emotion expression intensities in depression.
\newblock {\em Journal of Affective Disorders}, 2022.

\bibitem[\protect\citeauthoryear{Yin \bgroup \em et al.\egroup }{2022}]{yin2022intelligent}
W.~Yin, C.~Yu, P.~Wu, W.~Jiang, Y.~Liu, T.~Ren, and W.~Dai.
\newblock An intelligent mobile system for monitoring relapse of depression.
\newblock In {\em CSCW}, 2022.

\bibitem[\protect\citeauthoryear{Yoon \bgroup \em et al.\egroup }{2022}]{yoon2022d}
J.~Yoon, C.~Kang, S.~Kim, and J.~Han.
\newblock D-vlog: Multimodal vlog dataset for depression detection.
\newblock In {\em Proceedings of the AAAI Conference on Artificial Intelligence}, 2022.

\bibitem[\protect\citeauthoryear{Yuan \bgroup \em et al.\egroup }{2022}]{10011731}
X.~Yuan, S.~Zhang, C.~Zhao, X.~He, B.~Ouyang, and S.~Yang.
\newblock Pain intensity recognition from masked facial expressions using swin-transformer.
\newblock In {\em ROBIO}, 2022.

\bibitem[\protect\citeauthoryear{Zhang \bgroup \em et al.\egroup }{2023}]{Zhang2023DepressionRB}
F.~Zhang, M.~Wang, J.~Qin, Y.~Zhao, X.~Sun, and W.~Wen.
\newblock Depression recognition based on electrocardiogram.
\newblock {\em ICCCS}, 2023.

\bibitem[\protect\citeauthoryear{Zhao and Wang}{2022}]{zhao2022unaligned}
Z.~Zhao and K.~Wang.
\newblock Unaligned multimodal sequences for depression assessment from speech.
\newblock In {\em EMBC}, 2022.

\bibitem[\protect\citeauthoryear{Zhou \bgroup \em et al.\egroup }{2020}]{zhou2020visually}
X.~Zhou, K.~Jin, Y.~Shang, and G.~Guo.
\newblock Visually interpretable representation learning for depression recognition from facial images.
\newblock {\em IEEE Transactions on Affective Computing}, 2020.

\bibitem[\protect\citeauthoryear{Zhu \bgroup \em et al.\egroup }{2019}]{8653893}
J.~Zhu, Y.~Wang, R.~La, J.~Zhan, J.~Niu, S.~Zeng, and X.~Hu.
\newblock Multimodal mild depression recognition based on eeg-em synchronization acquisition network.
\newblock {\em IEEE Access}, 2019.

\end{thebibliography}
\end{spacing}
\end{document}